\documentclass[]{aastex631}

\usepackage[utf8]{inputenc}
\usepackage[T1]{fontenc}

\begin{document}

\title{Helium-burning blue large-amplitude pulsators: A Population Study with BPASS}

\email{wangbo@ynao.ac.cn}
\email{zhangzhengyang@ynao.ac.cn}

\author[0009-0006-8370-3108]{Zhengyang Zhang}
\affiliation{Yunnan Observatories, Chinese Academy of Sciences, Kunming 650011, PR China}
\affiliation{University of the Chinese Academy of Sciences, 19A Yuquan Road, Shijingshan District, Beijing 100049, PR China}
\affiliation{International Centre of Supernovae, Yunnan Key Laboratory, Kunming, 650216, PR China}

\author[0000-0002-6853-4055]{Conor M. Byrne}
\affiliation{Department of Physics, University of Warwick, Gibbet Hill Road, Coventry CV4 7AL, UK}

\author[0000-0002-2452-551X]{Chengyuan Wu}
\affiliation{Yunnan Observatories, Chinese Academy of Sciences, Kunming 650011, PR China}
\affiliation{International Centre of Supernovae, Yunnan Key Laboratory, Kunming, 650216, PR China}

\author[0000-0002-3231-1167]{Bo Wang}
\affiliation{Yunnan Observatories, Chinese Academy of Sciences, Kunming 650011, PR China}
\affiliation{International Centre of Supernovae, Yunnan Key Laboratory, Kunming, 650216, PR China}



\begin{abstract}
Blue Large-Amplitude Pulsators (BLAPs) are a class of radially pulsating stars with effective temperatures ranging from 20,000 to 35,000 K and pulsation periods between 7 and 75 minutes. This study utilizes the Binary Population and Spectral Synthesis (BPASS) code to investigate helium-burning stars as a formation channel for BLAPs in the Milky Way. The progenitor stars have initial masses of 3--6 $\rm M_{\odot}$, resulting in BLAPs with final masses of 0.5--1.2 $\rm M_{\odot}$. Based on a constant star formation rate of 3 $\rm M_{\odot} \,\rm yr^{-1}$ and solar metallicity (Z = 0.020), population synthesis predicts approximately 14,351 helium-burning BLAPs in the Milky Way: 12,799 with Main Sequence (MS) companions and 1,551 with evolved/compact-object companions. Helium-burning BLAPs show prolonged lifetimes in the pulsation region and a narrow stellar age range for entering this regime (log(t/yr) = 8.0--8.6), unlike pre-white dwarf models. BLAPs with MS companions typically form via Roche lobe overflow, leading to longer orbital periods ($\sim$100 days). Those with evolved/compact-object companions form through common envelope evolution, resulting in shorter periods. While Galactic extinction makes most BLAPs faint (apparent magnitudes $>$ 25), future surveys like WFST and VRO LSST are expected to detect approximately 500–900. This research establishes helium-burning stars as a significant BLAP contributor and offers testable predictions regarding their binary properties and Galactic distribution.
\end{abstract}



\section{Introduction} \label{sec1}
Blue Large-Amplitude Pulsators (BLAPs) were first discovered by \citet{2013AcA....63..379P, 2017NatAs...1E.166P} through the Optical Gravitational Lensing Experiment (OGLE) project. In recent years, approximately 200 BLAPs have been discovered, but only three BLAPs have been confirmed to reside in binary systems \citep{Pigulski2022, 2023NatAs...7..223L, 2024arXiv241208903Z, 2025arXiv250318379K,2025arXiv250708372P}. BLAPs are generally considered to be radially pulsating stars with effective temperatures ranging from 20,000 to 35,000 K and pulsation periods between 7 and 75 minutes. On the Hertzsprung-Russell Diagram, they are located between hot subdwarfs and hot Main Sequence(MS) stars. While BLAPs have effective temperatures similar to sdB stars, their surface gravity is about one order of magnitude lower. The log(g) for BLAPs typically ranges between approximately 4.2 and 5.7 dex.

BLAPs exhibit pulsation through the $\kappa$-mechanism, which is triggered by an increase in iron-peak element abundances (primarily $\rm ^{56}Fe$ and $\rm ^{58}Ni$) in their outer layers (about $\rm 10^{-10}\, M_{\sun}$). This mechanism also leads to the pulsational instability of sdB variables, which are characterized by effective temperatures similar to those estimated for BLAPs \citep[e.g.][]{1997ApJ...483L.123C, 2003ApJ...597..518F, 2006MNRAS.371..659J}. The generation of these enhanced elemental abundances is primarily attributed to atomic diffusion and radiative levitation processes \citep[e.g.][]{2011MNRAS.418..195H}. Notably, the incorporation of radiative levitation significantly expands the instability region \citep{2018MNRAS.481.3810B, 2020MNRAS.492..232B}.

Despite the discovery of about 200 BLAPs, our understanding of their formation remains incomplete. It is widely accepted that BLAPs are the stripped cores of evolved stars, but a low-mass star evolving in isolation cannot shed the requisite amount of its hydrogen-rich envelope within the age of the Universe. Consequently, their formation is thought to be dominated by binary interaction channels, such as Common Envelope ejection (CEE) or stable Roche lobe overflow (RLOF) \citep{2017NatAs...1E.166P, 2021MNRAS.507..621B}. Theoretical investigations have explored several possible outcomes of these binary interactions, leading to three main models for the internal structure of a BLAP. The first model proposes that BLAPs are low-mass (${\sim}0.3 \, \rm M_{\sun}$) pre-white dwarfs (pre-WDs) with a helium core supported by a hydrogen-burning shell \citep{2018MNRAS.477L..30R, 2020MNRAS.492..232B}. The second model suggests BLAPs are more massive stars undergoing core helium burning, placing them on an evolutionary track toward the extended horizontal branch \citep{2018MNRAS.478.3871W, 2018MNRAS.481.3810B, 2020ApJ...903..100M}. The third model posits they are in a subsequent stable helium shell burning phase, evolving away from the extended horizontal branch \citep{2022A&A...668A.112X, 2023NatAs...7..223L}.

Additionally, several merger scenarios have been suggested. The merger of a helium white dwarf (HeWD) with a low-mass main-sequence star (MS) could potentially form BLAPs \citep{2023ApJ...959...24Z}. Similarly, mergers of double extremely low-mass (DELM) white dwarfs may also produce BLAPs, particularly when the total system mass ranges from 0.32 to 0.7$\, \rm M_{\sun}$ \citep{2024A&A...691A.103K}. Particularly interesting is the possibility that BLAPs might be magnetic white dwarfs, which further emphasizes the importance of merger models in explaining their formation \citep{2024A&A...691A.343P}. BLAPs have now also been proposed to potentially originate from mergers of inner binary stars within triple or quadruple star systems, with BLAPs coexisting with companion stars having orbital periods of about 4000 days \citep{2025arXiv250318379K}. Nonlinear pulsation calculations are now also being used to verify the pulsation properties of BLAPs \citep{2024arXiv240816912J, 2025MNRAS.tmp..382J}.

Due to the currently limited discovery of BLAPs, their population in the Milky Way remains poorly understood. While rapid stellar population synthesis codes like BSE, COMPASS, and COSMIC can quickly generate evolutionary tracks for $10^{6}$ stars, they lack detailed elemental abundance information \citep{2000MNRAS.315..543H,2002MNRAS.329..897H,2020ApJ...898...71B,2022ApJS..258...34R}. BLAPs are specifically stars burning hydrogen in shells, helium in the core, or helium in shells. Without elemental composition data, determining their precise evolutionary state becomes challenging. Detailed stellar population synthesis codes like BPASS are therefore extremely valuable, as they provide comprehensive information including helium core mass, carbon core mass, helium abundance, and carbon abundance. These parameters allow for strict constraints on BLAP evolutionary state and formation channel. \citet{2021MNRAS.507..621B} used BPASS to implement stringent selection criteria for investigating potential pre-WD evolutionary channels. Their analysis revealed that BLAP progenitors strongly favor initial masses between 1 and 2$\, \rm M_{\sun}$. The initial orbital period ranges from approximately 1 to 400 days, with a peak of around 6 days, and encompasses a wide range of initial mass ratios. By analyzing the time each model spends in the BLAP region and the corresponding stellar population synthesis weights, they estimated that approximately 12,000 BLAPs exist in the Milky Way.

Although several formation channels for BLAPs have been proposed, the role of stars undergoing either core helium burning or shell helium burning has not been investigated from a population synthesis perspective. To address this gap, we employ BPASS to explore helium-burning stars as potential BLAP progenitors. Our analysis covers several key aspects: identifying viable formation pathways, estimating the expected number of such BLAPs in the Milky Way, and determining their characteristic distributions in terms of initial masses, companion properties, and orbital configurations. We extend our investigation to examine how these populations respond to varying metallicities and star formation histories. We found that helium-burning stars contribute significantly to the BLAP population, comparable to the previously recognized pre-WD channel. This finding suggests that the contribution of the helium-burning star channel to the BLAP population cannot be overlooked, emphasizing the need for further research into helium-burning scenarios.

The subsequent sections of this paper will detail our methodological approach in Section \ref{sec2}, present the key results derived from BPASS population synthesis simulations in Section \ref{sec3} and further discuss the implications of these findings, including the Galactic distribution of BLAPs and the impact of varying metallicities and star formation histories on these populations in Section \ref{sec4}, and finally, summarize our conclusions in the Section \ref{sec5}.

\section{Method}\label{sec2}
\subsection{BPASS}
The Binary Population and Spectral Synthesis (BPASS) code is designed to investigate the impact of massive binary stars on the observed spectra of young stellar populations at both solar and sub-solar metallicities \citep{2009MNRAS.400.1019E, 2018MNRAS.479...75S}. Additionally, \citet{2021MNRAS.507..621B} has utilized BPASS to explore low-mass stellar populations, such as pre-WD models of BLAPs. BPASS is based on a custom stellar evolution modeling code, which is derived from the STARS stellar evolution code \citep{1971MNRAS.151..351E,1995MNRAS.274..964P,2004MNRAS.353...87E, 2017PASA...34...58E}. This code tracks the evolutionary histories of individual stars and interacting binary systems, accounting for the effects of mass and angular momentum transfer to compute their structure, temperature, and luminosity. The BPASS code models all key binary interaction processes, including RLOF, CEE, and mergers, as this is a core function of a binary evolution code. To distinguish between the channels that produce our He-burning BLAP candidates, we differentiate between stable RLOF and CEE by examining the peak mass-transfer rate in the models, a method consistent with our approach in \citet{2021MNRAS.507..621B}. Merger events are identified directly via a specific model type flag in the BPASS output.

In the BPASS, detailed calculations are first performed to track the primary star, defined as the initially more massive star in the binary system. After determining the final evolutionary state of the primary star, the evolution of the secondary star (the initially less massive star) is replaced with a detailed model. This is done either by using a single-star model or by computing a new detailed binary model that includes a compact remnant, depending on whether the binary system remains bound after the primary star's evolution. Throughout this paper, we distinguish "primary stars," which refer to cases where the BLAP evolves from the primary star and is typically accompanied by a MS companion, and "secondary stars," which originate from the secondary star and may have an evolved/compact object companion.

The primary mass calculations cover a range from 0.1 to 300$\, \rm M_{\sun}$ in the parameter settings. The binary mass ratio is sampled uniformly between 0.1 and 0.9 with intervals of 0.1. The initial orbital period is sampled logarithmically from 0 to 4 (in log days) with steps of 0.2. BPASS assumes all binaries are in circular orbits (e=0). The main justification for this is the computational effort involved; adding eccentricity as an additional dimension to the model grid would have vastly increased the number of models required. For systems containing compact objects as companions, the model grid is determined based on the population synthesis results following the first supernova explosion. BPASS utilizes a modified Kroupa Initial Mass Function (IMF), which features a power-law index of -1.30 for stellar masses between 0.1 and 0.5$\, \rm M_{\sun}$, and -2.35 for masses up to 300$\, \rm M_{\sun}$ \citep{1993MNRAS.262..545K}.

Population synthesis is achieved through a weighted combination of primary stars. The weights are normalized such that the total stellar mass of the population is $\rm 10^{6} \, M_{\sun}$. These weights are determined by the joint probability distribution $W_{\rm pop}$, as detailed in \citet{2021MNRAS.507..621B}.

\subsection{Model Selection}
To investigate the contribution of helium-burning stars to the population of BLAPs, we make use of the detailed stellar evolution model data produced in BPASS, which tabulates stellar properties at discrete timesteps for each model. This data was accessed and manipulated using the Python package Hoki, providing data tables of stellar parameters as a function of time for every binary and single-star model \citep{2020JOSS....5.1987S}. We will first examine models with a metallicity of $Z = 0.020$, but we also consider other metallicities in subsequent sections.

We first exclude models where mass transfer begins during the AGB phase, using the criterion $M_{\rm carbon}/M > 0.01$ at the onset of mass transfer.
Subsequently, we establish five conditions to screen for helium-burning BLAPs (including core and shell helium burning) followed by the list:
\begin{enumerate}
    \item $4.4 < \log(T_{\rm eff}(\rm K)) < 4.55$
    \item $4 < \log(g/\text{cm s}^{-2}) < 6$
    \item $0.5 < M/\rm M_{\odot} < 1.2$
    \item $\text{Age} < 1.4 \times 10^{10}~\text{yr}$
    \item Helium mass is decreasing
\end{enumerate}

Condition 1 is established based on the maximum error in the effective temperature of OW-BLAP-3. Given that the maximum effective temperature of OW-BLAP-3 can reach 33,400 K, we choose an effective temperature close to the highest currently known for BLAPs \citep{2022MNRAS.513.2215R}. Condition 2 is established according to observational empirical data collected thus far. For condition 3, the lower limit corresponds to the upper mass bound from \citet{2021MNRAS.507..621B}, and the upper limit is set based on \citet{2018MNRAS.478.3871W}. Condition 4 must ensure the stellar age is less than the Universe's age. Condition 5 requires a decreasing helium abundance as a constraint on the helium-burning state and applies only when the first four conditions are met. The distinction between being in the core helium burning phase or the helium shell burning phase is made based on the decrease in the helium mass fraction and whether the helium core mass is zero.

Importantly, this selection depends on observed values and does not imply that all stars passing through this region will necessarily exhibit pulsational instability. The pulsation-driving mechanism for BLAPs arises from the opacity peaks of $\rm ^{56}Fe$ and $\rm ^{58}Ni$. If a star evolves through our selected region faster than the diffusion timescale, radiative levitation may not accumulate sufficient $\rm ^{56}Fe$ and $\rm ^{58}Ni$, potentially suppressing pulsation. A comprehensive pulsational stability analysis for the specific stellar structures predicted by helium-burning BLAP models is currently lacking. Such calculations are essential to accurately define the boundaries of their instability region, a task we will undertake in a forthcoming study.
The constraints outlined above may shift based on calculations of the pulsational instability region for helium-burning stars, thereby altering the number of models meeting the selection criteria. Based on observational data, the current selection provides a lower limit on the number of BLAPs.

\subsection{Population Synthesis}

Stellar population synthesis in BPASS relies primarily on the joint probability distribution $W_{\rm pop}$. Following a method similar to that of \citet{2021MNRAS.507..621B}, we assume a constant star formation rate (SFR) of $\rm 3\, M_{\sun}\, yr^{-1}$. Different star formation histories will be explored in the Discussion section. Given the SFR and the lifetime ($\Delta t$) that a model spends in the BLAP region, we can calculate the total mass of BLAPs ($M_{\rm BLAPs}$) produced by that model:
\begin{equation}
M_{\text{BLAPs}} = \text{SFR} \times \Delta t.
\end{equation}
Using the equation modified from \citet{2021MNRAS.507..621B}:
\begin{equation}
N_{\text{stars}} = \frac{M_{\text{BLAPs}} \times W_{\text{pop}}}{(1 + q)M_{\text{init}}}.
\end{equation}
We calculate the number of BLAPs ($N_{\rm stars}$) for this model, where $q$ represents the initial mass ratio and $M_{\rm init}$ is the initial mass of the primary star. By applying this procedure to each model, we can calculate the number of BLAPs across all models. These values yield the total number of BLAPs in the population.

\section{Result}\label{sec3}

\subsection{Selection result}
After applying our selection criteria, we obtained 209 models. Of the 209 models, 165 are primary stars and 44 are secondary stars. Classified by burning state, there are 203 core helium-burning models and 6 shell helium-burning models. Among the primary BLAPs, one is a shell helium-burning star, while there are five among the secondary BLAPs. We present a Kiel diagram of these models in Figure \ref{fig1}, accompanied by observed BLAP data (see Appendix \ref{table1} for full details). We classify the observed BLAPs into two groups based on $\log N({\rm He})/N({\rm H})$ from the observational sample: one group with values greater than -2 and another with values less than or equal to -2 (This is equivalent to a dividing line at [He/H] = -0.93). Helium-burning star models can account for both low-gravity and high-gravity BLAPs. Low-mass helium-burning stars, upon entering the BLAP region, have a smaller radius and exhibit a higher $\log g$ compared to more massive helium-burning stars, before subsequently evolving toward higher effective temperatures. Moreover, helium abundance cannot serve as a criterion to distinguish between high-gravity and low-gravity BLAPs.

\begin{figure}[ht!]
\plotone{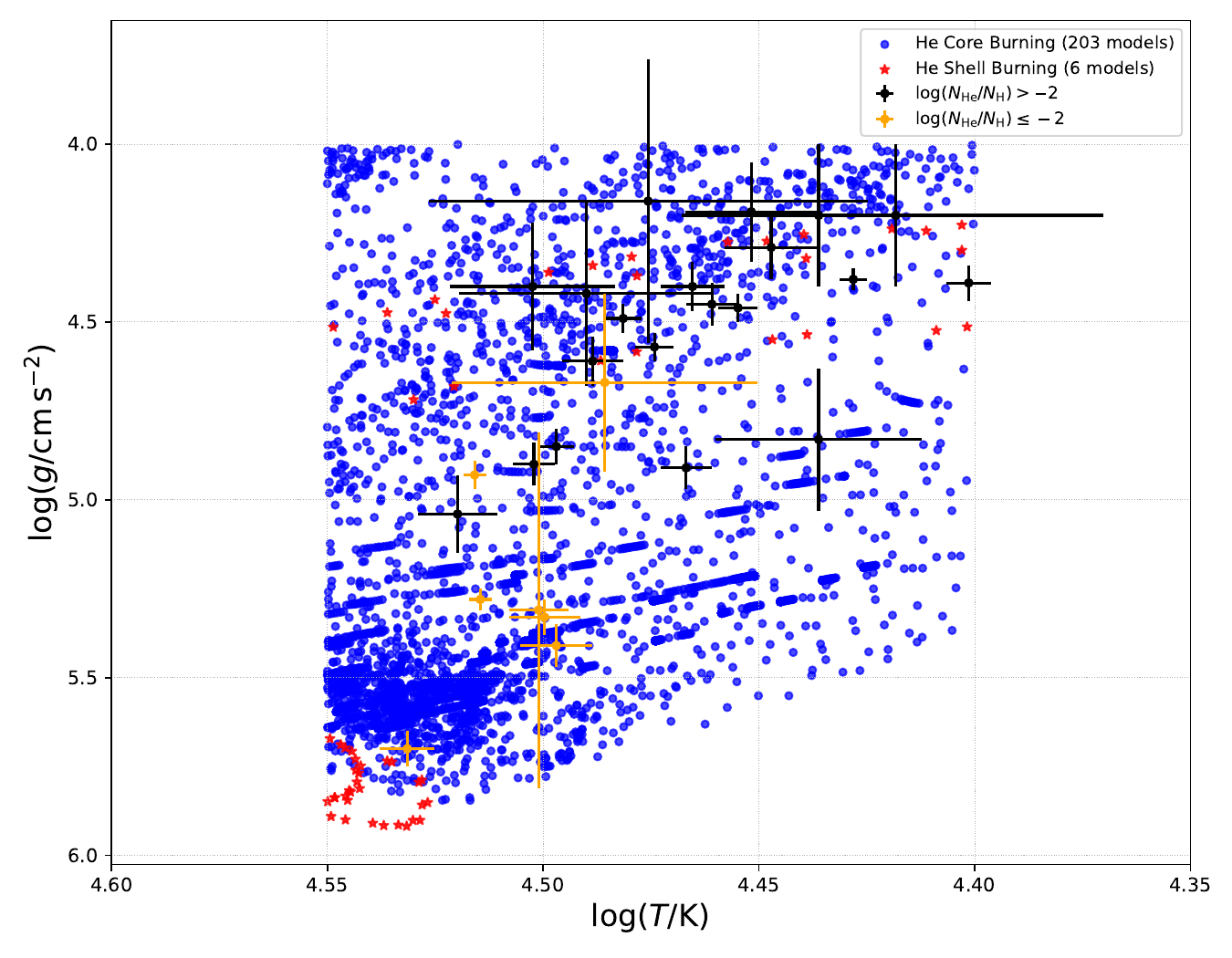}
\caption{The Kiel diagram displaying the effective temperature versus surface gravity of helium-burning BLAP candidates. Observed BLAPs are shown as crosses: black crosses denote objects with $\log (N_{\rm He}/N_{\rm H}) > -2$, while orange plus symbols denote objects with $\log (N_{\rm He}/N_{\rm H}) \le -2$ He Core Burning models are indicated by blue circles (203 models), and He Shell Burning models are indicated by red asterisks (6 models).
\label{fig1}}
\end{figure}

\begin{figure}[ht!]
\plotone{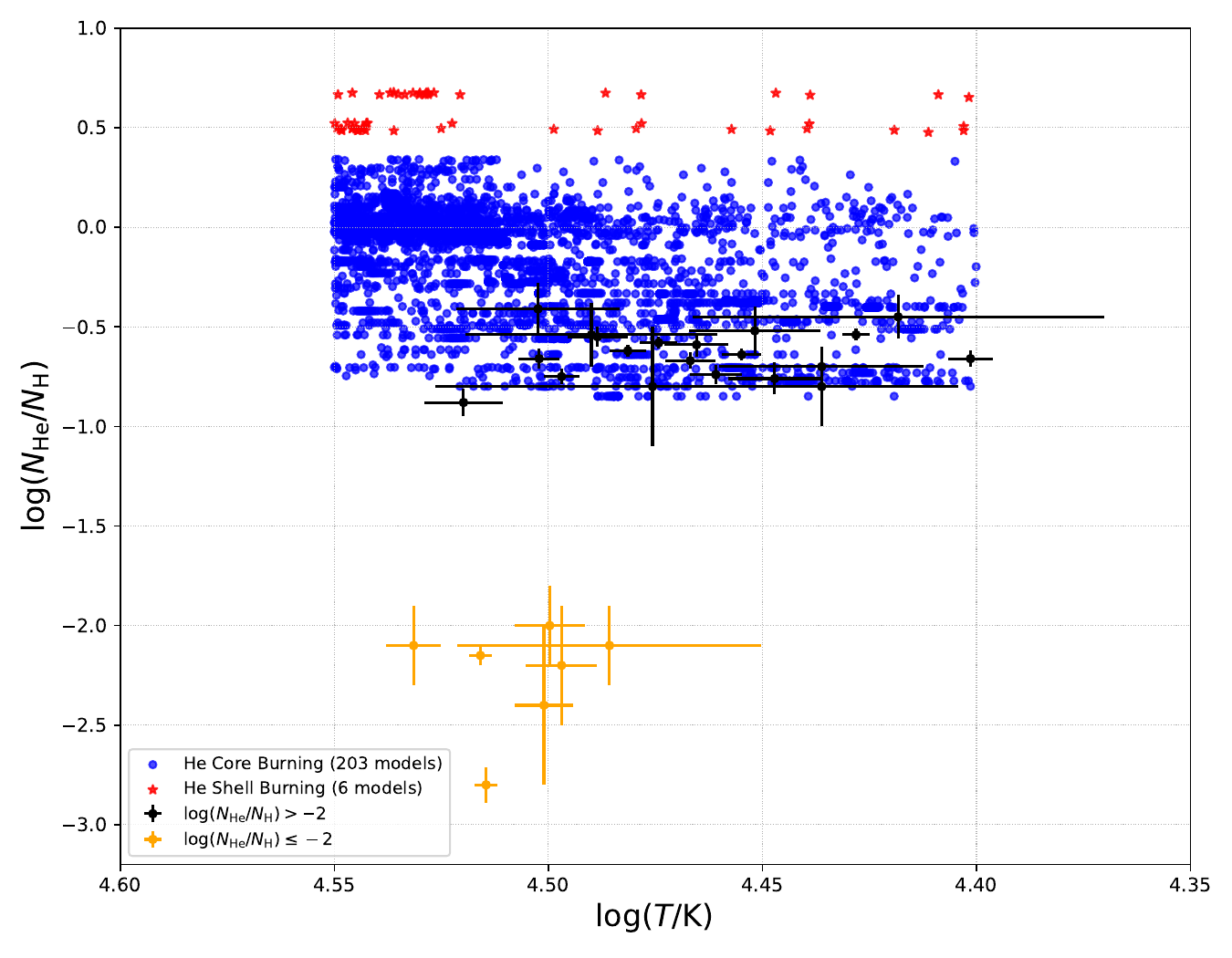}
\caption{Helium abundance [He/H] versus effective temperature log$(T/\rm K)$. Blue dots represent our 203 core helium burning models, while red asterisks show the 6 shell helium burning models. Observed BLAPs are plotted with error bars and are distinguished by helium content: black crosses denote helium-rich objects with $\log (N_{\rm He}/N_{\rm H}) > -2$, and orange plus signs indicate helium-poor objects with $\log (N_{\rm He}/N_{\rm H}) \leq -2$.
\label{fig2}}
\end{figure}

We also present a comparison between the model's $\log (N_{\rm He}/N_{\rm H})$ and observational data in Figure \ref{fig2}. It is found that the model can only account for BLAPs with $\log (N_{\rm He}/N_{\rm H}) > -1$. However, helium-deficient BLAPs cannot be explained by either the helium-burning star model or the pre-WD model \citep{2021MNRAS.507..621B}. Interestingly, the surface helium abundance of shell helium-burning models is significantly higher than that of core helium-burning models. Nevertheless, we note that the model by \citet{2023ApJ...959...24Z} successfully explains both helium-rich and helium-poor BLAPs. This suggests that helium-poor BLAPs are more likely to originate from the HeWD-MS merger channel. It is also possible that the inability to explain helium-poor BLAPs stems from the lack of detailed consideration of the element diffusion process in the BPASS calculations.

\begin{figure}[ht!]
\plotone{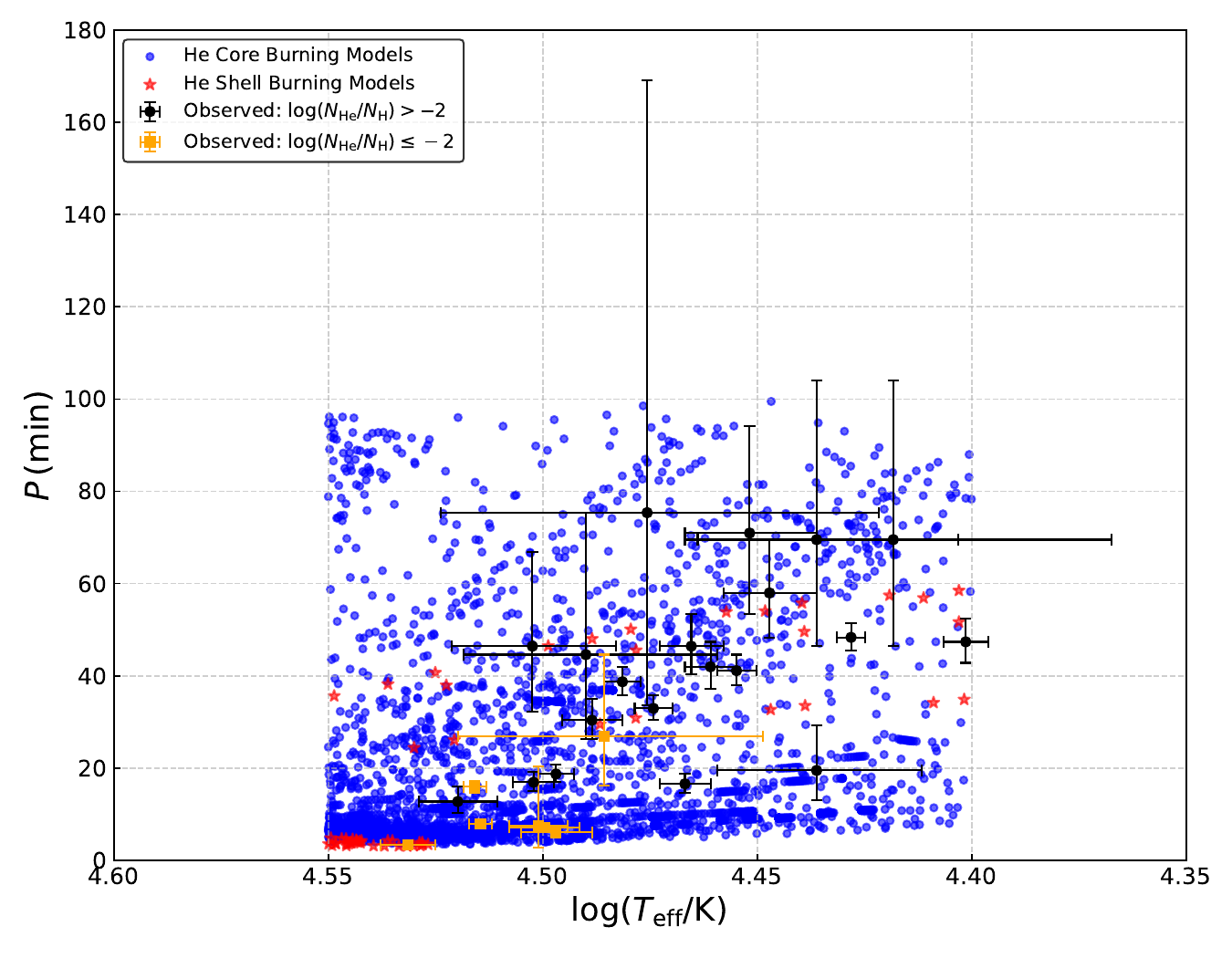}
\caption{The distribution of pulsation periods as a function of effective temperature for our model predictions and observed BLAPs. Our theoretical models are represented by blue dots for core helium burning models and red asterisks for shell helium burning models. Observed BLAPs are plotted with error bars and are distinguished by helium abundance: black crosses denote helium-rich objects with $\log (N_{\rm He}/N_{\rm H}) > -2 $, while orange plus signs indicate helium-poor objects with $\log (N_{\rm He}/N_{\rm H}) \le -2 $.
\label{fig3}}
\end{figure}

We also made a rough estimate of whether the pulsation periods of the model match the observational data. BLAPs are generally considered to be radial fundamental-mode pulsators, so the mean density-period relation can effectively estimate their pulsation periods. The formula we use is:
\begin{equation}P\approx Q_\mathrm{F}/\sqrt{\frac{M}{M_\odot}\left(\frac{R_\odot}{R}\right)^3}\mathrm{~minutes},\end{equation}
where $Q_\mathrm{F} \approx 47 \, \text{min}$ \citep[e.g.][]{10.1093/mnras/79.1.2, 2023ApJ...959...24Z}.
Figure \ref{fig3} shows that for both the core and shell helium burning phases, the pulsation periods of our model BLAPs are within the observed range and consistent with the data. Furthermore, the pulsation period distribution predicted by our model effectively encompasses the observed distribution.

\begin{figure}[ht!]
\plotone{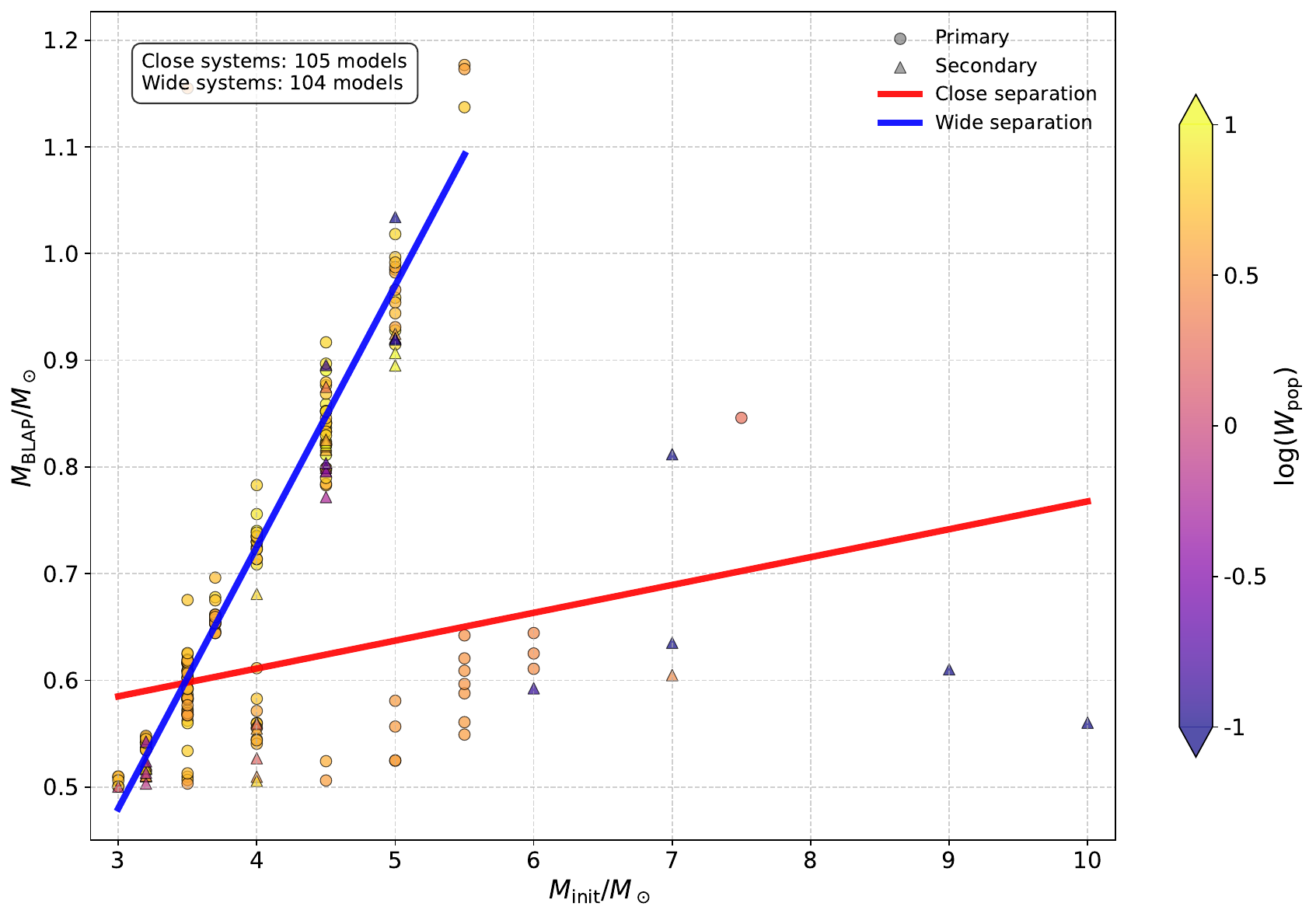}
\caption{The relationship between initial stellar mass ($M_{\text{init}}$) and final BLAP mass ($M_{\text{BLAP}}$) for our binary evolution models. The models separate into two distinct populations based on their initial binary separation. A linear regression fit is shown for each group: a shallow trend for close separation systems (red line) and a much steeper, stronger correlation for wide separation systems (blue line). The color of each point corresponds to its population synthesis weight ($\log W_{\text{pop}}$), as indicated by the color bar.
\label{fig4}}
\end{figure}

Figure~\ref{fig4} illustrates the relationship between the initial mass ($M_{\text{init}}$) and the final BLAP mass ($M_{\text{BLAP}}$). The initial masses of our models are predominantly concentrated between $3-6\, M_{\odot}$, a range consistent with the findings of \citet{2018MNRAS.478.3871W}. A key finding from our analysis is that this relationship is not monolithic; instead, it bifurcates into two distinct trends based on the initial binary separation. Systems with a wide separation exhibit a strong, positive correlation (blue line), where the final BLAP mass increases steeply with the initial mass. In contrast, systems with a close separation show a much weaker, nearly flat correlation (red line), with final BLAP masses clustered around $0.5-0.6\, M_{\odot}$ largely independent of the initial mass. This bifurcation can be explained by the timing of the mass transfer. In close systems, the short initial orbital period causes stable RLOF to begin early in the star's post-main sequence evolution. This early and efficient mass transfer strips away almost the entire hydrogen envelope, consistently leaving behind a helium core of a relatively fixed mass, which explains the weak correlation. Conversely, in wider systems, mass transfer begins later when the progenitor has already developed a more massive helium core, leading to the strong positive correlation between initial and final mass. The color of each model point indicates its population synthesis weight ($\log W_{\text{pop}}$). The number of primary stars significantly exceeds that of secondary stars, and the primary stars also have a higher weight. Consequently, in subsequent population synthesis studies, primary stars are expected to be more abundant in the Galaxy. It is noteworthy that secondary BLAPs appear in cases with initial masses of $9\,\rm M_\odot$ and $10\,\rm M_\odot$.  In these two models, their companions are compact objects with initial masses of $3.16\,\rm M_\odot$ and $3.98\,\rm M_\odot$ respectively.  We deduce that rare BLAP-black hole binary systems will form in these scenarios. Such exceptions are infrequent in stellar populations and are generally negligible. However, the number of systems with a secondary BLAP and a white dwarf/neutron star (BLAP-WD/NS) companion is on the order of $10^2$ to $10^3$.

\begin{figure}[ht!]
\centering
\includegraphics[width=1.0\columnwidth]{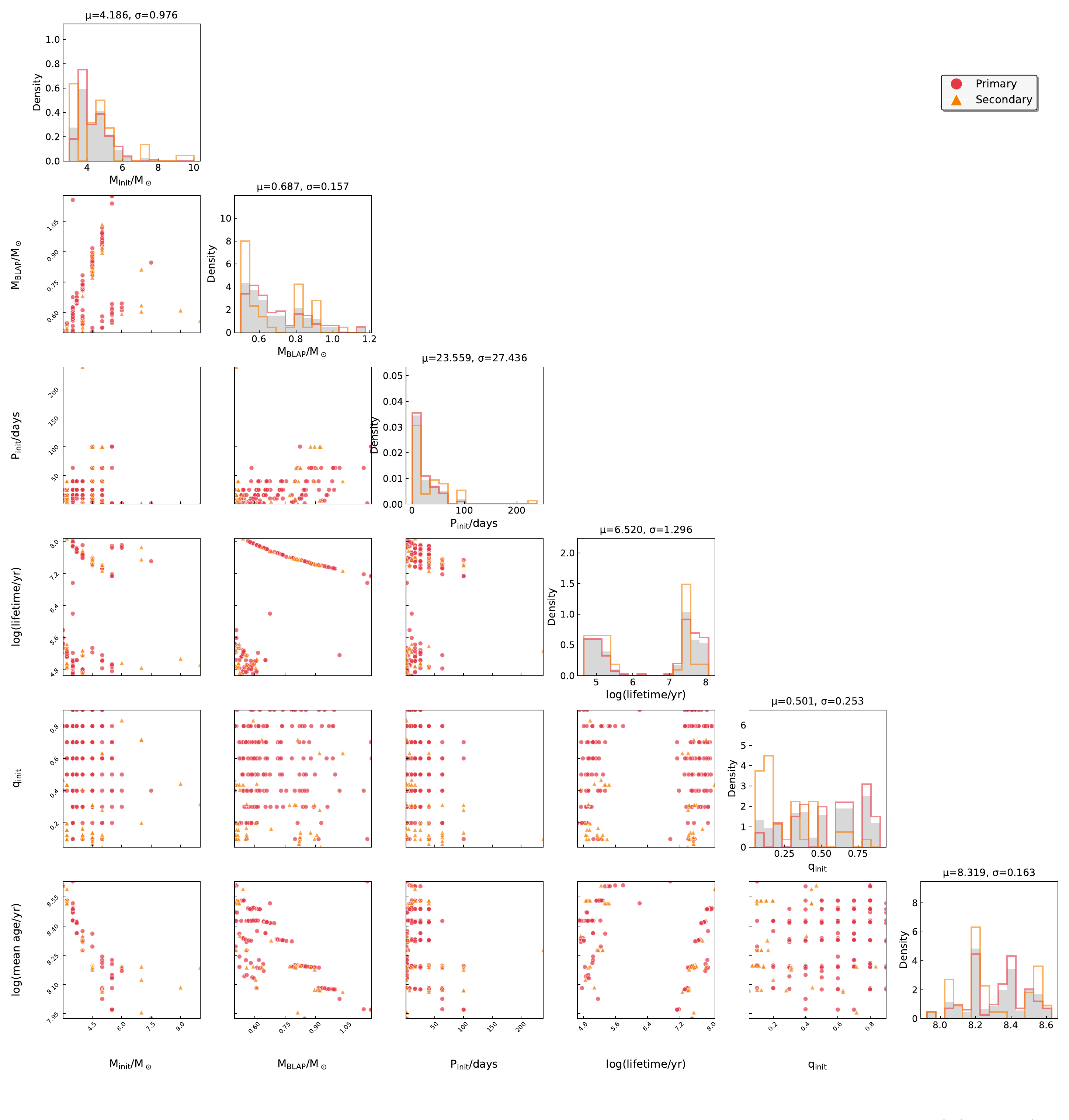}
\caption{A corner plot illustrating the distributions and pairwise correlations for key parameters of our simulated BLAPs. The parameters shown are the initial mass ($M_{\text{init}}$), final BLAP mass ($M_{\text{BLAP}}$), orbital period ($P_{\text{orb}}$), BLAP lifetime, initial mass ratio ($q_{\text{init}}$) and age. 
The diagonal panels display the 1D histograms for each parameter, annotated with their mean ($\mu$) and standard deviation ($\sigma$). The off-diagonal panels show the 2D scatter plots for each parameter pair. 
The models are separated into primary BLAPs (red dots) and secondary BLAPs (orange triangles), with their corresponding distributions in the histograms shown by the red and orange outlines, respectively. The filled gray histogram in the background represents the combined statistical distribution of all models together, serving as a reference for the overall trends.
\label{fig5}}
\end{figure}

To explore the properties of the progenitor systems that form BLAPs, we present a corner plot in Figure~\ref{fig5}. This figure displays the one-dimensional distributions and two-dimensional correlations for six key parameters: initial mass ($M_{\text{init}}$), final BLAP mass ($M_{\text{BLAP}}$), orbital period ($P_{\text{orb}}$), BLAP lifetime(The duration of the BLAP phase of a star), initial mass ratio ($q_{\text{init}}$), and age. The models are separated into primary and secondary BLAPs to highlight differences in their formation channels.

The diagonal panels reveal the statistical distributions of these parameters. The initial mass ($M_{\text{init}}$) distribution peaks around $4\,M_{\odot}$, while the final BLAP masses ($M_{\text{BLAP}}$) are typically clustered around a mean of $\mu \approx 0.69\, M_{\odot}$. The initial orbital periods are heavily concentrated at short values ($P_{\text{orb}} < 50$\,days). An interesting feature is seen in the initial mass ratio ($q_{\text{init}}$) distribution: while primary BLAPs are dominant overall, secondary BLAPs become a significant fraction of the population at low mass ratios ($q_{\text{init}} \lesssim 0.4$), suggesting they tend to form in systems with large initial mass differences.

The off-diagonal panels uncover key physical correlations. A clear negative correlation exists between the final BLAP mass and age, consistent with stellar evolution theory, where more massive stars have shorter main-sequence lifetimes. Perhaps the most significant finding is the bimodal distribution of the BLAP lifetime, with one population of models having lifetimes around $10^{7.5}\,$yr and another, shorter-lived population clustered around $10^{5.5}$\,yr. The number of models in the long-lived group is the largest, and they have a longer duration in the BLAP stage. Although the number of models is not as large as that of the pre-white dwarf models, the advantage in lifespan also indicates that they should constitute a significant portion of the observed BLAP population.

We can offer a tentative explanation for this bimodal lifetime distribution. In the long-lifetime region, stars enter the BLAP phase at approximately 18.5\% of their total age. With a moderate helium core fraction, they maintain stable helium burning and follow the scaling relation $t \sim M^{-2.8}$, which leads to a more extended evolution within the BLAP region. Conversely, in the short-lifetime region, stars enter the BLAP phase much later, at around 71.2\% of their age. By this point, helium burning is in its final stages and may deviate from the scaling relation. Because they are at the end of their helium-burning phase, these stars evolve out of the BLAP region much more quickly.

\subsection{Population}
\begin{figure}[ht!]
\centering
\includegraphics[width=0.6\textwidth]{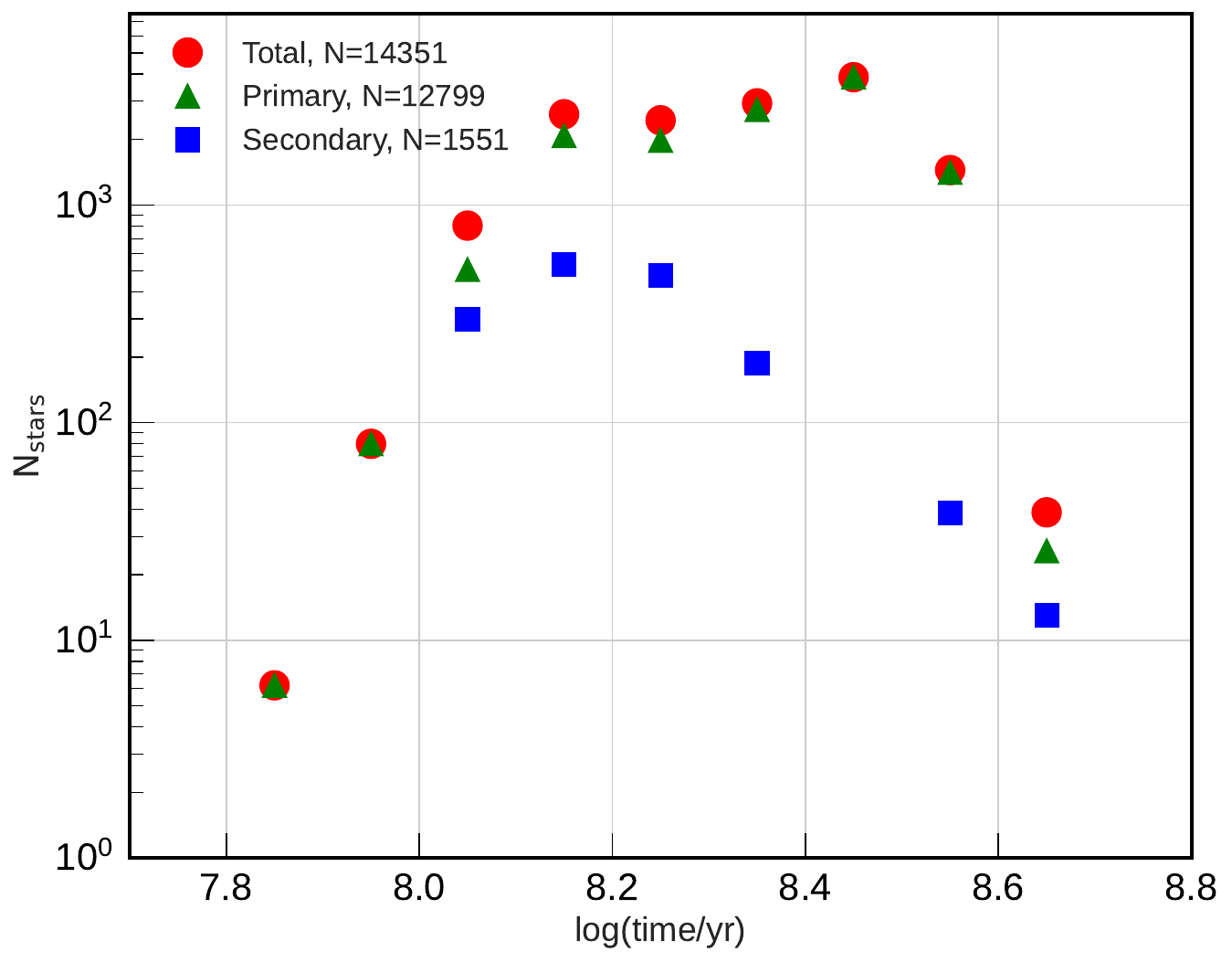}
\caption{Temporal distribution of BLAPs in the Milky Way based on our population synthesis calculations. The total number of BLAPs (red circle) is approximately 14,351, comprising 12,799 primary stars (green triangles) and 1,551 secondary stars (blue squares).
\label{fig6}}
\end{figure}

If the distribution across time bins of BLAPs is not considered, our calculations indicate that there are 14,373 BLAPs in the Milky Way, including 12,832 primary stars and 1,541 secondary stars. However, when accounting for the time bins evolution of the BLAP population, there may be regions with rapid changes in number where insufficient sampling points could lead to underestimation or overestimation of their contribution. To address this, we employ a high-precision numerical integration method to calculate the number of BLAPs accurately. Specifically, we uniformly sampled 1,000 points on a logarithmic time scale, covering the range from the earliest evolutionary entry time to the longest evolutionary lifetime (\(\log(t/\text{yr}) =\) 6-10.2). For each time point, we calculated the instantaneous contribution rate of all models. We numerically integrated the instantaneous contribution rates to obtain the total number of BLAPs. Considering the precision requirements, we used Simpson's integration method. As shown in Figure \ref{fig6}, the total number of BLAPs in the Milky Way is approximately 14,351, including 12,799 primary stars and 1,551 secondary stars. This result is very close to the theoretical expectation of 14,373. Additionally, we found that compared to pre-WDs, the time distribution of helium-burning BLAPs is more concentrated, primarily within the range of \(\log(t/\text{yr}) = 8.0\) to 8.6. This distribution is quite expected, as helium-burning stars have larger MS masses and shorter evolutionary timescales.

\begin{figure}[ht!]
\plotone{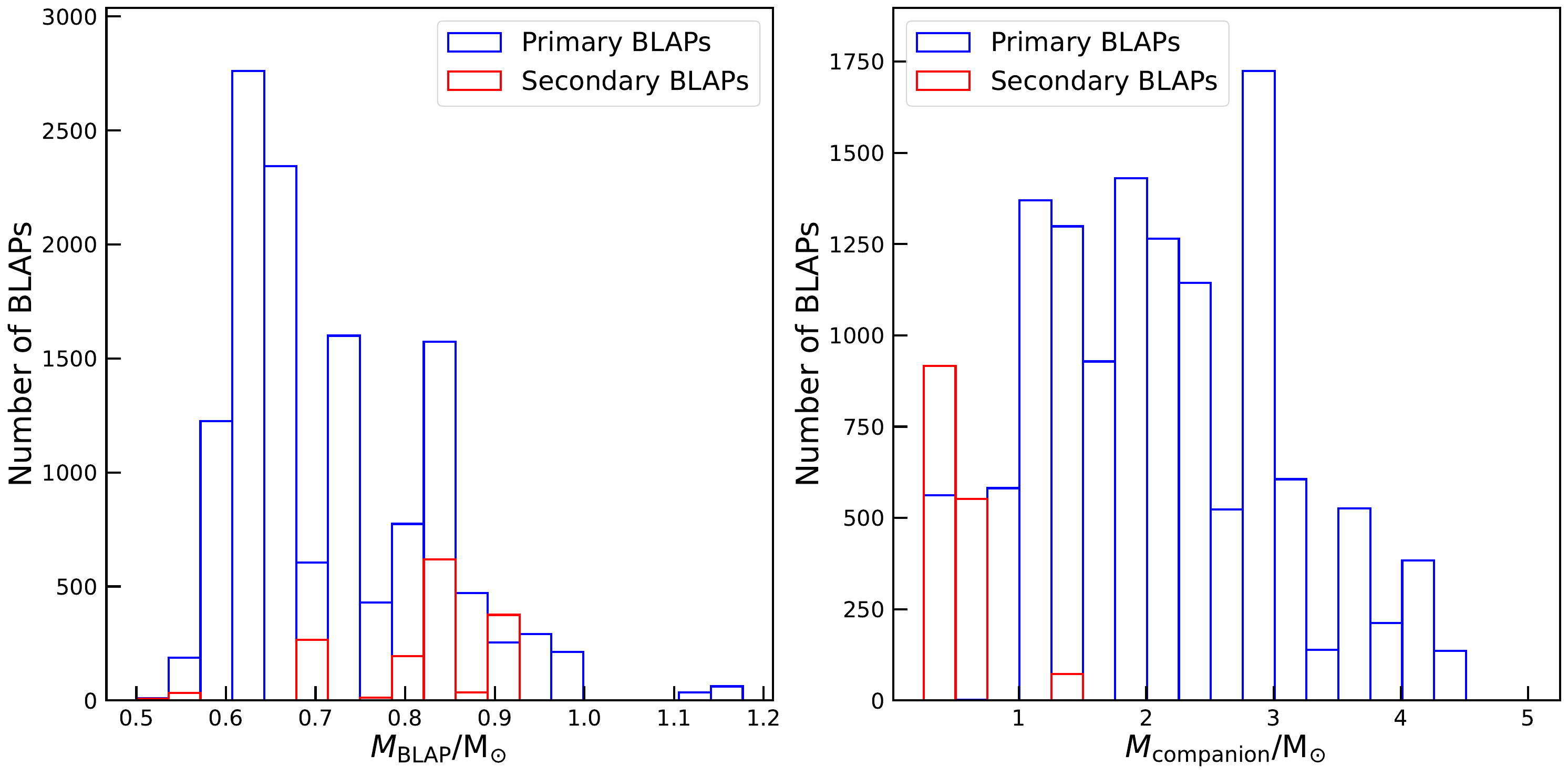}
\caption{Mass distributions for BLAPs and their companions. The left panel shows a histogram of masses for primary (blue) and secondary (red) BLAPs. The right panel shows the mass distribution of the corresponding companion stars in these binary systems.
\label{fig7}}
\end{figure}

Figure \ref{fig7} illustrates the mass distributions of Blue Large-Amplitude Pulsators (BLAPs) and their companion stars. In the left panel, we observe a prominent mass peak for primary BLAPs within the range of approximately $0.6 - 0.7 \, \rm M_{\sun}$, while secondary BLAPs exhibit a distinct peak near approximately $0.83 \, \rm M_{\sun}$. The right panel presents the mass distribution of companion stars, spanning a wide range from $0.2 \, \rm M_{\sun}$ to $5 \, \rm M_{\sun}$. The companions of primary BLAPs show a significant peak at around $2.8 \, \rm M_{\sun}$, suggesting a preference for relatively massive MS companions. According to \citet{2022A&A...668A.112X}, for most spectral types of MS companions, the flux of the BLAP dominates; however, sufficiently hot main-sequence stars (e.g., A-type or F-type) may dominate in the optical or infrared bands, depending on the stellar radius. In unresolved binary systems, only the total flux can be observed. The companion’s flux alters the shape of the spectral energy distribution, particularly at longer wavelengths (e.g., mid-infrared), where an M-type companion contributes an excess of approximately 0.1 magnitudes, with earlier spectral types producing a substantially larger effect. Variations in the effective temperature and radius of a BLAP during its pulsation cycle may further complicate the detection of its companion. Although \citet{Pigulski2022} discovered that the MS companion mass of a binary BLAP is $2.8 \rm M_{\sun}$, additional observational examples are still needed to validate the mass distribution of companions.

In contrast, the companions of secondary BLAPs predominantly have masses below $1 \, \rm M_{\sun}$, strongly indicating that these are WDs. Although a few models produce companions with masses greater than $2\,\rm M_{\sun}$, their population-weighted contribution is negligible (totaling only 0.65) due to their very low synthesis weights. The prevalence of WD companions carries significant observational implications: given that white dwarfs are intrinsically faint and challenging to detect, this may partly account for the scarcity of observed BLAP binary systems in current surveys.

\begin{figure}[ht!]
\plotone{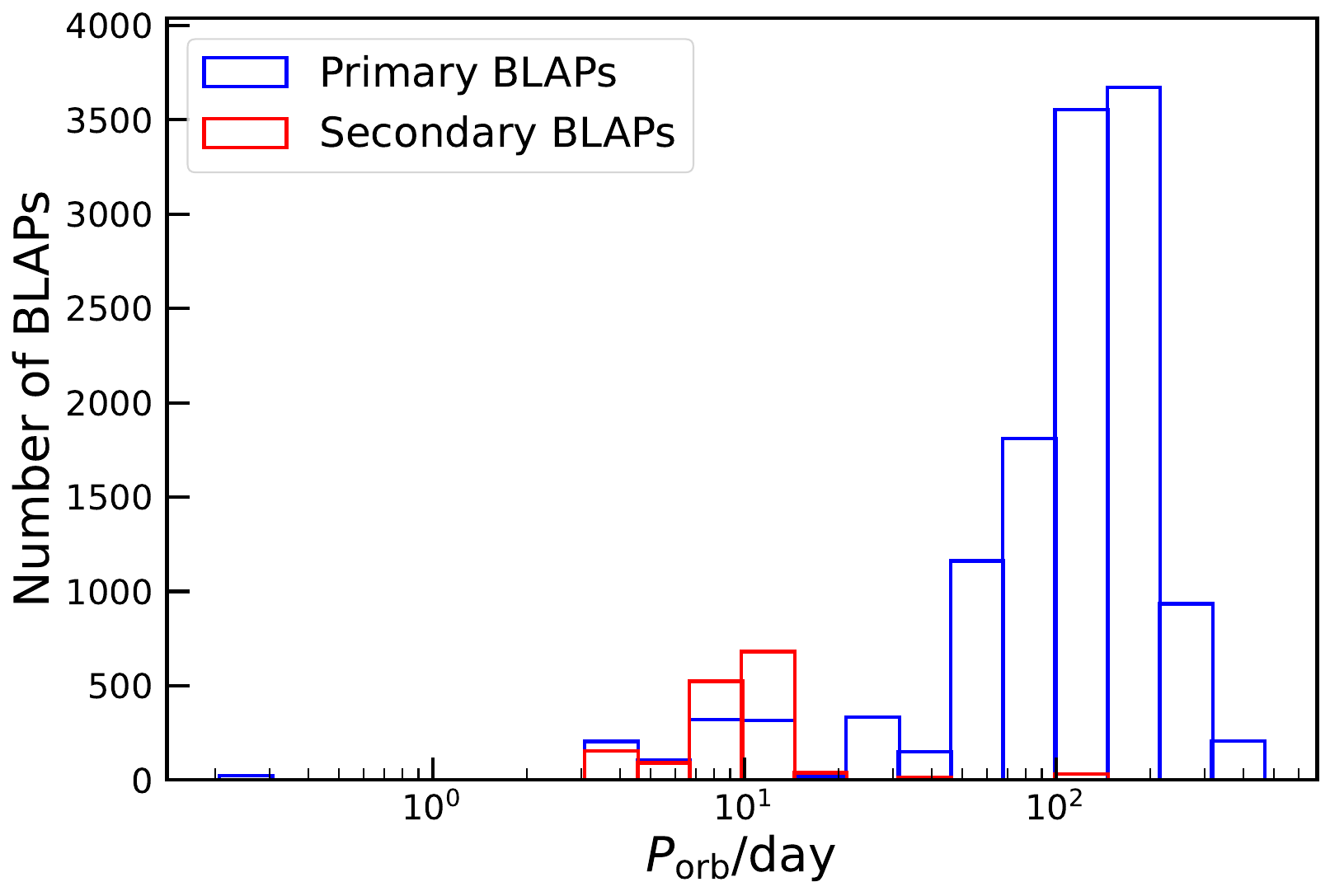}
\caption{Comparison of orbital period distributions for primary and secondary BLAPs. The plot reveals two distinct populations: secondary BLAPs (red) have short periods clustering around 10 days, while primary BLAPs (blue) have significantly longer periods, peaking above 100 days. 
\label{fig8}}
\end{figure}

Figure \ref{fig8} illustrates the orbital period distribution of primary and secondary BLAPs in binary systems, revealing a clear dichotomy between the two groups. Primary BLAPs predominantly occupy longer-period systems (around 100 days), suggesting that these systems likely evolved through RLOF. This evolutionary pathway typically maintains wider orbits while facilitating the mass transfer necessary for BLAP formation. Although the extended orbital periods pose observational challenges, these may be mitigated by analyzing archival satellite data from long-term monitored systems, following the example of the first confirmed BLAP binary system. In contrast, secondary BLAPs display a mix of short and long orbital periods in comparable numbers, indicating that similar fractions of these BLAPs have undergone either CE evolution or RLOF. However, the proportion of secondary BLAPs shaped by CE evolution significantly exceeds that of primary BLAPs. The CE phase, characterized by rapid in-spiral of the binary components, naturally results in shorter final orbital periods. Thus, this orbital period distribution provides compelling evidence for distinct dominant formation channels: RLOF primarily governs the formation of primary BLAPs, while CE evolution plays a more significant role for secondary BLAPs.

\section{Discussion}\label{sec4}
\subsection{Galactic BLAP distribution}
To date, all known BLAPs have been found near the Galactic plane. With the exception of the brightest, relatively nearby object HD133729, all BLAPs are located within a band of $\left | b \right |<12^{\circ}$ \citep[e.g.][]{2025arXiv250708372P}. For a simplified estimation, we modeled the three-dimensional stellar number density distribution in the disk using a double exponential profile, expressed as:
\begin{equation}\rho_d(r,z,M)=\exp\left(-\frac{r-r_0}{h}\right)\exp\left(-\frac{|z|}{H(M)}\right)\end{equation}

where \(\rho_d\) denotes the stellar number density of the disk component. The first exponential, $\exp(-(r - r_0)/h)$, is the radial decay term, describing the decrease in stellar density moving outwards from the Galactic center. In this term, $r$ is the radial distance from the Galactic center (The range is from about 0 to 20kpc), $r_0$ is the distance from the Sun to the Galactic center (assumed to be 8 kpc), and $h$ is the disk scale length. For this work, we adopt a value of $h = 3$ kpc.

The second exponential is the vertical decay term, which models the thin structure of the disk. Here, $z$ is the perpendicular distance from the Galactic plane (The range is from about -1.4kpc to 1.4kpc). The parameter $H(M)$ is the exponential scale height. For our simplified estimation, we adopt a constant average scale height of $H(M) = 300$ pc for the BLAP population \citep{1980ApJS...44...73B}.

To distribute the BLAPs within this disk, we assumed a random distribution weighted by the overall stellar density \(\rho_d\), ensuring that BLAPs are more likely to appear in regions of higher stellar density, consistent with the general stellar population in the disk. Extinction is also a critical factor in our analysis. We employed the S-model from \citet{2005AJ....130..659A}, which uniquely accounts for the Milky Way’s spiral arm structure. Consequently, extinction exhibits step-like variations with increasing distance.

\begin{figure}[ht!]
\centering 
\includegraphics[width=1.0\textwidth]{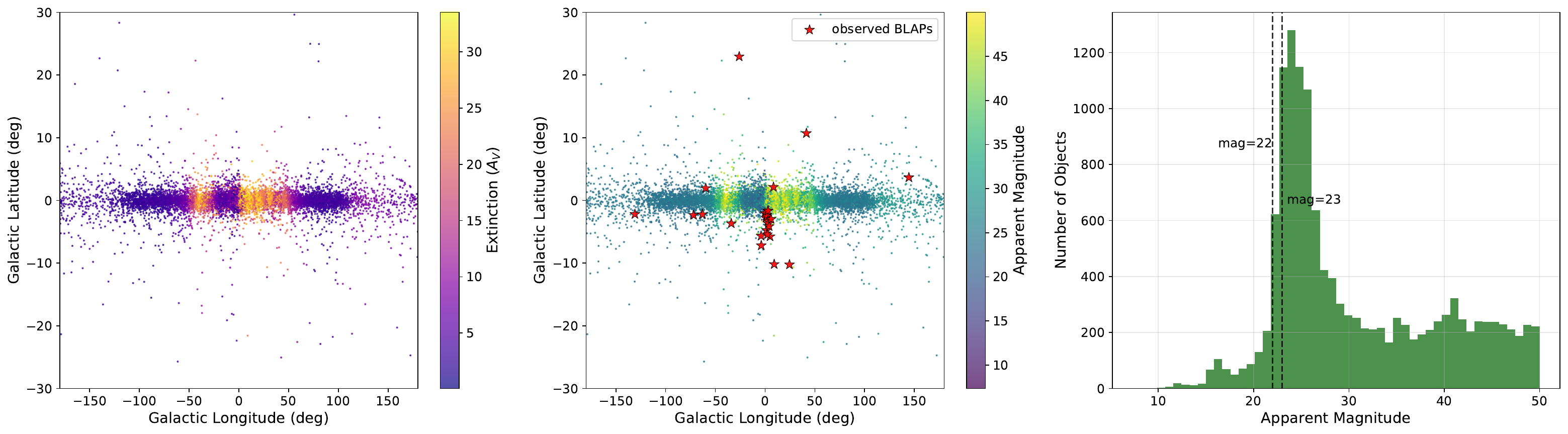}
\caption{A three-panel overview of the simulated BLAP population. Left panel: A 2D map of the interstellar extinction model projected onto Galactic coordinates. Center panel: A projection of the simulated BLAP population onto the sky (Galactic longitude and latitude). The color of each point represents its calculated apparent magnitude. The positions of known, observationally confirmed BLAPs are overlaid as red stars for comparison. Right panel: A histogram showing the apparent magnitude distribution for the entire simulated population. The dashed vertical lines indicate typical magnitude limits for astronomical surveys, providing a prediction for the number of discoverable objects.
}
\label{fig9}
\end{figure}

Figure \ref{fig9} presents a comprehensive overview of our BLAP model, from its physical inputs to its final observable predictions. The figure is composed of three panels that illustrate the simulation workflow and results. The left panel displays a two-dimensional map of the interstellar extinction model used in our simulation, projected onto the sky in Galactic coordinates (l,b). The color scale represents the extinction magnitude. The extinction is most severe in the direction of the Galactic center, near a longitude of $l \approx 0^\circ - 30^\circ$. This pronounced increase is a result of the tangential direction effect, where the line of sight runs parallel to the spiral arm, accumulating a massive column density of dust. This detailed extinction map is a critical input for our simulation, as it directly affects the calculated apparent magnitudes of the synthetic BLAP population.

The middle panel illustrates the projected distribution of our simulated BLAP population in the Galactic coordinate system ($l, b$), where each point is colored by its calculated apparent magnitude. Crucially, we have overlaid the positions of currently known, observationally confirmed BLAPs (marked as red stars) to provide a direct comparison between our model's predictions and real-world data. The g-band absolute magnitude of a BLAP is simply estimated as 4 mag \citep{2018A&A...620L...9R}. A striking feature of our simulation is the predicted deficit of detectable sources in the direction of the Galactic center ($l \approx 0^\circ$). This is a direct consequence of the high-density interstellar dust in this region, which causes the apparent magnitudes of most BLAPs to become substantially fainter, pushing them beyond the limits of typical surveys. Interestingly, despite this high extinction, a significant fraction of the known BLAPs have been discovered by targeted, high-cadence surveys such as OGLE precisely in these challenging, dust-obscured regions. This success in the Galactic center leads to the exciting prospect that current and future all-sky, high-cadence surveys will be highly effective at discovering a large new population of BLAPs, and potentially BLAP binary systems, in the less extincted directions away from the Galactic plane.

The right panel depicts the apparent magnitude distribution of the predicted BLAP population, which peaks around magnitude 25, indicating that most BLAPs lie beyond current observational limits. We evaluated two key detection thresholds. The Wide Field Survey Telescope (WFST), a northern sky survey instrument with a 30-second exposure time, achieves a 5$\sigma$ detection limiting magnitude of approximately 22 (g-band) \citep[e.g.][]{2023SCPMA..6609512W, 2023RAA....23c5013L, 2024arXiv241212601L}. Located in the Northern Hemisphere, it is well-suited for observing the outer regions of the Milky Way and the Galactic disk. Based on this limiting magnitude, we estimate that approximately 500 BLAPs are detectable. In contrast, the Vera C. Rubin Observatory’s Legacy Survey of Space and Time (VRO LSST) reaches a deeper limiting magnitude of $g = 23$, corresponding to a 5\% photometric uncertainty in single-epoch observations \citep[e.g.][]{2009arXiv0912.0201L}. This increased sensitivity is expected to raise the number of detectable BLAPs to approximately 900. However, a secondary consideration arises: at these limiting magnitudes, detecting statistically significant brightness variations—characteristic of BLAPs—may be challenging, particularly for the faintest objects. With single-epoch photometric uncertainties approaching or exceeding typical BLAP amplitudes (0.2--0.4 mag) near the detection threshold, confirming variability could require extensive observations. Thus, while VRO LSST’s depth and cadence enhance its potential, the estimate of 900 detectable BLAPs likely represents an optimistic upper limit, with the actual number of identifiable pulsators potentially lower due to these observational constraints. We also investigated the number of helium-burning BLAPs potentially observable by the OGLE project. The OGLE survey's depth in the $g$-band reaches an apparent magnitude of 19. Consequently, the estimated number of BLAPs visible within the OGLE survey region is about 7. However, when accounting for the intrinsic luminosity variations of BLAPs, this number is likely to be even smaller. Currently, the number of BLAPs detected by OGLE is nearly 100. However, only one, OGLE-BLAP-009, has evidence of being a helium-burning star. Because our model for the Milky Way is very simplistic—consisting only of an exponential disk—our estimate of the total population is inevitably biased.

We acknowledge that the uncertainty in the extinction model is impactful, especially considering the large number of BLAPs predicted to be near the limiting magnitude of surveys. For example, in a survey with a detection limit of 22 mag, an object with a calculated apparent magnitude of 21.9 mag would have a true magnitude of 21.9 $\pm$ 0.25 mag, considering the rms uncertainty reported by \citet{2005AJ....130..659A}. This implies that a fraction of such objects may in reality be fainter than 22 mag and would therefore be missed. While a full re-analysis is beyond the scope of the current work, future studies could incorporate this uncertainty using a Monte Carlo method. Such an approach would allow for a more precise quantification of the impact on the overall detection efficiency and survey yield.

These simulation results demonstrate the significant impact of interstellar extinction on observational data. We suggest that when designing survey projects targeting BLAPs, careful consideration must be given to the extinction properties of the target sky regions. In particular, it is advisable to avoid directions near the Galactic center at low Galactic latitudes and areas that spiral arms may obscure. The choice of observational band is also critical. BLAPs are relatively bright in the optical due to their high temperature (with peak emission in the blue-optical range). However, a BLAP with a $T_{\rm eff}$ of ~30,000 K emits less in the near-infrared (the Rayleigh-Jeans tail), resulting in a fainter absolute magnitude in those bands. The optimal trade-off, therefore, likely lies in the red-optical bands, such as SDSS r (623 nm) or i (763 nm), where the effect of extinction is moderately reduced while the BLAP flux remains high.

\subsection{Different metallicities and star formation rates}
\citet{2015ApJ...808..132H}, using observational data from 69,919 red giant stars in the SDSS-III/APOGEE survey, examined the distribution of stellar chemical abundances across various regions of the Galactic disk, revealing significant variations in chemical composition. Near the Galactic plane (\(|z| < 0.5\) kpc), [Fe/H] displays a distinct radial gradient: in the inner disk (\(3 < R < 5\) kpc), the [Fe/H] distribution peaks at approximately +0.32; in the solar neighborhood (\(7 < R < 9\) kpc), it peaks at about +0.02; and in the outer disk (\(13 < R < 15\) kpc), it peaks around -0.48. Given this broad range of metallicities, investigating the population of BLAPs across different metallicity regimes is crucial. However, at low metallicities, it remains uncertain whether radiative levitation can concentrate sufficient iron and nickel to create an opacity peak capable of driving pulsations. Sustaining pulsations under such conditions may be challenging. Thus, the discussion of low-metallicity environments here is presented merely as a hypothetical scenario.

\begin{figure}[ht!]
\centering
\includegraphics[width=0.6\textwidth]{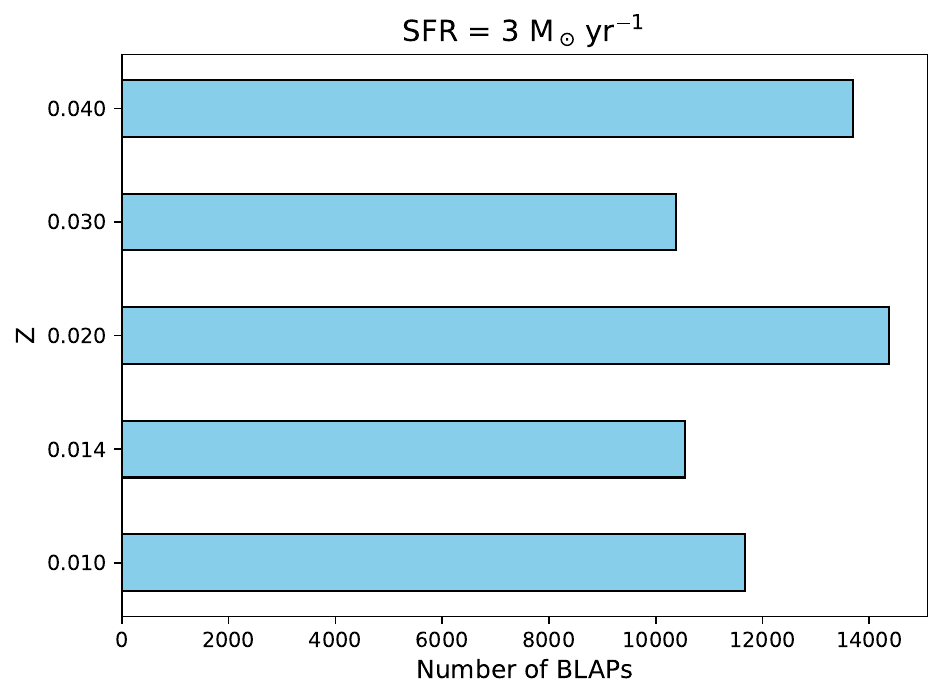}
\caption{Number of helium-burning BLAPs in the Milky Way (x-axis) as a function of metallicity (y-axis), derived from BPASS population synthesis simulations assuming a constant star formation rate of \(\mathrm{SFR} = 3 \, \mathrm{M}_{\odot} \, \mathrm{yr}^{-1}\). 
\label{fig10}}
\end{figure}

In our population synthesis simulations using BPASS, we investigated the impact of metallicity on the predicted number of BLAPs during their helium-burning phase in the Milky Way, assuming a constant star formation rate of 3 \(\rm M_\odot \, yr^{-1}\). For metallicities \( Z = 0.010 \), 0.014, 0.020, 0.030, and 0.040, the corresponding BLAP counts are 11,664, 10,556, 14,373, 10,372, and 13,695, respectively (see Figure \ref{fig10}). These results indicate that the BLAP population peaks at near-solar metallicity (\( Z = 0.020 \)), with reduced numbers at both lower and higher metallicities. Notably, the counts at \( Z = 0.014 \) and \( Z = 0.030 \) are lower than those at adjacent metallicities, suggesting a complex, non-monotonic relationship between metallicity and BLAP formation efficiency. Recent studies reveal that the fraction of low-mass stars in young stellar populations increases with metallicity \citep[e.g.][]{2023Natur.613..460L}. However, this trend alone does not account for the non-monotonic variation in BLAP counts. Instead, this pattern likely arises from metallicity’s influence on stellar evolution pathways, mediated by effects on opacity, mass loss, and binary interaction processes—key factors in forming BLAP progenitors.

It is interesting to note that the general trend in total BLAP numbers with metallicity is roughly similar to the trends observed in \citet{2021MNRAS.507..621B}, with a peak at \( Z = 0.020 \). This peak suggests that near-solar metallicity conditions are most favorable for BLAP formation. Given the distinct radial metallicity gradient in the galaxy, as highlighted by \citet{2015ApJ...808..132H}, and the strong density gradient with radius (as outlined in our disk model in Section 4.1), it could be insightful to consider a simplistic model dividing the galaxy into radial intervals and distributing BLAPs with decreasing metallicity towards the outer regions. However, since the weighted mean metallicity across the galaxy is likely close to \( Z = 0.020 \), such a model would probably yield a BLAP population estimate similar to that obtained from assuming a uniform metallicity of \( Z = 0.020 \). 

The SFR is a crucial parameter for galaxies. Over the past few decades, various methods have been employed to estimate the Milky Way's SFR, including techniques such as Lyman Continuum Photon Rates, Massive Star Counts and Supernova Rates, among others \citep[e.g.][]{1978A&A....66...65S, 2005AJ....130.1652R}. These methods, after standardization to a common framework, yield largely consistent SFR estimates, converging to approximately $1.9 \pm 0.4 \rm M_{\sun} \, yr^{-1}$ \citep{2011AJ....142..197C}.

The star formation rate is a critical parameter in estimating the population of BLAPs in the Milky Way. Our initial calculations assumed a constant SFR of 3 $\rm M_\odot$ yr$^{-1}$ which yielded an estimated 14,351 helium-burning BLAPs at 
$Z = 0.020$. However, adopting a more realistic SFR for the Milky Way significantly reduces the BLAP population. Specifically, for $Z = 0.020$ and an SFR of 1.9 $\rm M_\odot$ yr$^{-1}$ the number of helium-burning BLAPs decreases to 9,103. This reduction is expected, as a lower SFR implies a decreased rate of stellar formation, directly affecting the number of progenitors available to evolve into BLAPs across all metallicities. The trend is consistent with the proportional relationship between SFR and BLAP counts, as fewer stars formed per unit time lead to fewer BLAPs under identical evolutionary conditions.

When considering metallicity-dependent star formation histories (SFHs), such as those modeled to account for variations near the solar neighborhood, we find that such SFHs are not necessary for simulating the populations of helium-burning BLAPs \citep[e.g.][]{2021MNRAS.501..302A}. In these SFHs, each coefficient \({\rm a}_{i}\) is proportional to the number of stars formed between \((t_i, t_i + dt)\) with metallicity between \((Z_j, Z_j + dZ)\). The SFR is derived by dividing \({\rm a}_{i}\) by the intervals \(dt\) and \(dZ\). The primary reason for the unsuitability is the temporal distribution of helium-burning BLAPs, which is concentrated within the first 1 Gyr of stellar evolution. According to the coefficients (\({\rm a}_{i}\)) from such SFHs, only for \( Z=0.030 \) is there a non-zero \({\rm a}_{i}\) value before 1 Gyr. Consequently, applying these SFHs would result in a negligible number of helium-burning BLAPs, rendering the results neither meaningful nor representative of the broader population.

In contrast, pre-WD BLAPs exhibit a broader temporal distribution, extending over periods where the \({\rm a}_{i}\) coefficients are significant. This broader distribution aligns better with the SFHs that include delayed or extended star formation epochs, making such models more applicable for estimating pre-WD BLAP populations. Therefore, while metallicity-dependent SFHs may be suitable for pre-WD BLAPs, they are not appropriate for helium-burning BLAPs due to the latter's concentrated evolutionary timescale.

\citet{2017MNRAS.469.4763M} proposed a novel single-degenerate model to explain Type Ia supernovae (SNe Ia), termed the Common Envelope Wind (CEW) model. According to their model, the surviving companion of a Type Ia supernova can evolve into a BLAP. However, the treatment of the common envelope process in BPASS differs from the CEW model \citep{2017PASA...34...58E}. Therefore, we do not consider this scenario for the surviving companion.

\subsection{Comparison with the Pre-White Dwarf Channel}
Now that synthetic stellar populations have been created for both the helium-burning channel (this work) and the pre-WD channel \citep{2021MNRAS.507..621B}, we can identify several key differences between their progenitors. The progenitors of helium-burning BLAPs primarily originate from intermediate-mass stars with initial masses of 3--6\,\( \rm M_{\odot} \), which is higher than the progenitors of pre-WD BLAPs that favor lower-mass stars of 1--3\,\( \rm M_{\odot} \). Furthermore, helium-burning BLAPs have a concentrated age distribution, reflecting the rapid evolution of their intermediate-mass progenitors. In contrast, pre-WD BLAPs span a much broader range of ages (up to several Gyr) due to the longer main-sequence lifetimes of their low-mass progenitors. Despite these differences, both channels are predicted to be significant contributors to the total BLAP population in the Milky Way. Assuming a constant star formation rate of $\rm 3\, M_{\sun}\, yr^{-1}$, population synthesis predicts approximately 14,351 helium-burning BLAPs. This is comparable to the estimated 12,000 BLAPs expected to form through the pre-WD channel.

\section{Conclusion}\label{sec5}
This study presents the first systematic exploration of helium-burning stars in Galactic binary systems as the progenitors of BLAPs, using the BPASS. Our findings establish that helium-burning stars constitute a significant formation channel for BLAPs, contributing an estimated 14,351 BLAPs to the Galactic population under a constant SFR of \( 3 \, {\rm M}_{\odot} \, \text{yr}^{-1} \) at solar metallicity (\( Z = 0.020 \)). This number is comparable to the approximately 12,000 BLAPs predicted from the previously dominant pre-WD channel \citep{2021MNRAS.507..621B}, highlighting that helium-burning pathways are equally critical to understanding BLAP evolution.

The helium-burning BLAP progenitors primarily originate from stars with initial masses between 3 and 6 \( \rm M_{\odot} \), evolving into BLAPs with final masses ranging from 0.5 to 1.2 \( \rm M_{\odot} \). Primary BLAPs dominate the population (12,799) compared to secondary BLAPs (1,551), reflecting their higher formation efficiency in binary systems. Companions to primary BLAPs are typically more massive (peaking around 2.8 \( \rm M_{\odot} \)), while secondary BLAPs are often paired with WDs (masses \( < 1 \, \rm M_{\odot} \)). It is noteworthy that in BPASS, the mass peak of the companion to the primary star closely aligns with that of the MS companion in the binary BLAP system HD 133729. The orbital period further distinguishes these two types: primary BLAPs tend to exhibit longer periods (approximately 100 days), suggesting formation through RLOF evolution, whereas secondary BLAPs display a short-period distribution, indicating that CE evolution plays a significant role in their formation.

Our population synthesis study indicates that the population of BLAPs peaks at solar metallicity ($Z=0.020$), with their numbers exhibiting a complex, non-monotonic dependence on metallicity and a proportional relationship with the SFR. We estimate that the OGLE survey can detect 7 BLAPs originating from this channel. Future survey projects, such as the WFST and the VRO LSST, are optimistically projected to detect approximately 500 and 900 BLAPs, respectively, in regions less affected by interstellar extinction. The BLAP magnitude distribution derived in this study suggests that the discrepancy between the population synthesis results and the observed scarcity of BLAPs can be partially alleviated. The lower observed number can be attributed to the fact that the vast majority of BLAPs are subject to severe interstellar extinction, causing their magnitudes to exceed the detection limits of current telescopes. Additionally, the limited sky coverage of OGLE and the severe extinction in the bulge region further contribute to this discrepancy.

This study demonstrates that the helium-burning and pre-WD channels appear to contribute comparably to the total predicted number of BLAPs in the Milky Way. The helium-burning BLAP model resides in the defined pulsational region for a longer duration than the pre-WD BLAP model, albeit with a smaller population number. Helium-burning BLAPs originate from stars with a narrower age distribution (between \( \log(t/\text{yr}) = 8.0 \) and 8.6), whereas pre-WD BLAPs exhibit a broader age distribution, suggesting potential differences in the progenitor populations and environments from which they arise. It is entirely plausible (and reasonable) that distinct BLAP formation pathways coexist, without necessitating a single pathway to account for all BLAPs.

\begin{acknowledgments}
We thank the anonymous reviewer for the valuable comments and suggestions. This study is supported by the National Natural Science Foundation of China (Nos 12288102, 12225304, 12090040/12090043, 12473032), the National Key R\&D Program of China (No. 2021YFA1600404), the science research grant from the China Manned Space Project (No. CMS-CSST-2021-A12), the Yunnan Revitalization Talent support Program (Yunling Scholar Project), the Yunnan Revitalization Talent Support Program (Young Talent project), the Yunnan Fundamental Research Project (No. 202201BC070003, 202501AS070005, 202501AW070001), the United Kingdom Science and Technology Facilities Council (STFC) Consolidated Grant (Number T/X001121/1) and the International Centre of Supernovae, Yunnan Key Laboratory (No. 202302AN360001).
\end{acknowledgments}
\bibliography{work3}{}
\bibliographystyle{aasjournal}

\appendix

\section{Table of BLAP parameters}

\begin{deluxetable*}{lccccccc}[h]
\tablewidth{0pt}
\label{table1}
\tablecaption{This appendix presents a compilation of parameters for previously known BLAPs, which are used for comparison in our study. }
\tablehead{
\colhead{Name} & \colhead{$T_{\rm eff}[\rm K]$} & \colhead{log $g$[dex]} & \colhead{log $N({\rm He})/N({\rm H})$} & \colhead{$l[^\circ]$} & \colhead{$b[^\circ]$} & \colhead{References}}
\startdata
OGLE-BLAP-001 & 30,800$\pm$500 & 4.61$\pm$0.07 & -0.55$\pm$0.05 & 288.06355 & -2.34712 & (1)\\ 
OGLE-BLAP-009 & 31,800$\pm$1400 & 4.40$\pm$0.18 & -0.41$\pm$0.13 &  2.90511 & -1.65832 & (1)\\
OGLE-BLAP-010 & 29,800$\pm$300 & 4.40$\pm$0.18 & -0.41$\pm$0.13 & 355.94188 & -5.66572 & (1)\\
OGLE-BLAP-019 & 28,000$\pm$700 & 4.29$\pm$0.09 & -0.76$\pm$0.08 & 359.94513 & -2.05015 & (1)\\
OGLE-BLAP-020 & 29,200$\pm$500 & 4.40$\pm$0.07 & -0.59$\pm$0.06 & 8.41333 & 2.12352 & (1)\\
OGLE-BLAP-021 & 28,500$\pm$300 & 4.46$\pm$0.04 & -0.64$\pm$0.03 & 0.82920 & -2.77380 & (1)\\
OGLE-BLAP-022 & 28,900$\pm$400 & 4.45$\pm$0.06 & -0.74$\pm$0.05 & 1.86967 & -2.26578 & (1)\\
OGLE-BLAP-024 & 25,200$\pm$300 & 4.39$\pm$0.05 & -0.66$\pm$0.04 & 1.73988 & -2.56094 & (1)\\
OGLE-BLAP-030 & 31,400$\pm$300 & 4.85$\pm$0.05 & -0.75$\pm$0.04 & 2.83324 & -3.26300 & (1)\\
OGLE-BLAP-031 & 26,800$\pm$200 & 4.38$\pm$0.03 & -0.54$\pm$0.03 & 356.10354 & -7.19029 & (1)\\
OGLE-BLAP-033 & 33,100$\pm$700 & 5.04$\pm$0.11 & -0.88$\pm$0.07 & 3.25208 & -3.68113 & (1)\\
OGLE-BLAP-034 & 30,300$\pm$300 & 4.49$\pm$0.04 & -0.62$\pm$0.03 & 1.43366 & -4.75695 & (1)\\
OGLE-BLAP-037 & 32,800$\pm$200 & 4.93$\pm$0.04 & -2.15$\pm$0.05 & 4.91343 & -5.76037 & (1)\\
OLGE-BLAP-042 & 28,300$\pm$1000 & 4.19$\pm$0.14 & -0.52$\pm$0.12 & 296.95618 & -2.22606 & (2)\\
OGLE-BLAP-044 & 32,700$\pm$200 & 5.28$\pm$0.03 & -2.80$\pm$0.09 & 300.26251 & 1.94534 & (2)\\
OGLE-BLAP-049 & 29,300$\pm$400 & 4.91$\pm$0.06 & -0.67$\pm$0.04 & 325.99337 & -3.67615 & (2)\\
OW-BLAP-1 & 30,600$\pm$2500 & 4.67$\pm$0.25 & -2.1$\pm$0.2 & 2.29558 & -5.43142 & (3)\\
OW-BLAP-2 & 27,300$\pm$1500 & 4.83$\pm$0.20 & -0.7$\pm$0.1 & 4.03441 & -4.13026 & (3)\\
OW-BLAP-3 & 29,900$\pm$3500 & 4.16$\pm$0.40 & -0.8$\pm$0.3 & 5.96087 & -2.98851 & (3)\\
OW-BLAP-4 & 27,300$\pm$2000 & 4.20$\pm$0.20 & -0.8$\pm$0.2 & 2.90514 & -1.65831 & (3)\\
TMTS-BLAP-1 & 31,780$\pm$350 & 4.90$\pm$0.06 & -0.66$\pm$0.05 & 144.52596 & 3.65689 & (4)\\
HD 133729 & 29,000$\pm$1800 & 4.5 & ... & 333.95658 & 22.93880 & (5)\\
High-gravity-BLAP-1 & 34,000$\pm$500 & 5.70$\pm$0.05 & -2.1$\pm$0.2 & 229.08074 & -2.19685 & (6)\\
High-gravity-BLAP-2 & 31,400$\pm$600 & 5.41$\pm$0.06 & -2.2$\pm$0.3 & 9.11465 & -10.18263 & (6)\\
High-gravity-BLAP-3 & 31,600$\pm$600 & 5.33$\pm$0.05 & -2.0$\pm$0.2 & 24.49480 & -10.23755 & (6)\\
High-gravity-BLAP-4 & 31,700$\pm$500 & 5.31$\pm$0.05 & -2.4$\pm$0.4 & 41.50915 & 10.71026 & (6)\\
\enddata
\tablecomments{
     For HD 133729, the log(g) was assumed by \citet{Pigulski2022} (see their Sect. 4).
}
\tablerefs{
    (1) \citet{2024arXiv240416089P}; 
    (2) \citet{2023AcA....73....1B}; 
    (3) \citet{2022MNRAS.513.2215R}; 
    (4) \citet{2023NatAs...7..223L}; 
    (5) \citet{Pigulski2022};
    (6) \citet{2019ApJ...878L..35K}.
}
\end{deluxetable*}

\end{document}